\documentclass[%
    reprint,
    amssymb,
    amsmath,
    aip,
    rsi,
]{revtex4-2}

\usepackage{amssymb}
\usepackage{graphicx}
\usepackage{docs}
\usepackage{bm}
\usepackage[colorlinks=true,linkcolor=blue]{hyperref}
\usepackage{amsmath}
\usepackage{siunitx}
\usepackage{float}
\DeclareSIUnit\amagat{amg}

\usepackage{todonotes}
\usepackage{tabularx, multirow}

\usepackage{physics}
\usepackage{color}
\usepackage{ulem}
\definecolor{myviolet}{rgb}{0.5,0,0.5} 
\definecolor{myred}{rgb}{1,0,0} 
\definecolor{myorange}{rgb}{1,.5,0} 
\definecolor{mygreen}{rgb}{0,0.5,0} 
\definecolor{myblue}{rgb}{0,0,0.75} 
\definecolor{mybrown}{rgb}{0.384314, 0.203922, 0.0705882} 
\definecolor{mymagenta}{cmyk}{0,1,0,0.12} 

\newcolumntype{Y}{>{\centering\arraybackslash}X}

\newcommand{\bn}{\va{n}}
\newcommand{\bx}{\va{x}}
\newcommand{\by}{\va{y}}
\newcommand{\bz}{\va{z}}
\newcommand{\bB}{\va{B}}
\newcommand{\bS}{\va{S}}
\newcommand{\etal}{\textit{et al.}}

\expandafter\ifx\csname package@font\endcsname\relax\else
\expandafter\expandafter
\expandafter\usepackage
\expandafter\expandafter
\expandafter{\csname package@font\endcsname}
\fi
\def \beq {\begin{equation}}
\def \eeq {\end{equation}}

\usepackage[english]{babel}
\addto\extrasenglish{%
\def\sectionautorefname~#1\null{Sec.\,#1\null}
\def\subsectionautorefname~#1\null{Sec.\,#1\null}
\def\subsubsectionautorefname~#1\null{Sec.\,#1\null}
\def\equationautorefname~#1\null{Eq.\,(#1)\null}
\def\figureautorefname~#1\null{Fig.\,#1\null}
\def\appendixautorefname~#1\null{Appendix \,#1\null}
} 

\begin{document}

\title{Meridional composite pulses for low-field magnetic resonance}

\author{Sven Bodenstedt}
\affiliation{ICFO -- Institut de Ci\`encies Fot\`oniques, The Barcelona Institute of Science and Technology, 08860 Castelldefels (Barcelona), Spain}

\author{Morgan W. Mitchell}
\affiliation{ICFO -- Institut de Ci\`encies Fot\`oniques, The Barcelona Institute of Science and Technology, 08860 Castelldefels (Barcelona), Spain}
\affiliation{ICREA -- Instituci\'{o} Catalana de Recerca i Estudis Avan\c{c}ats, 08010 Barcelona, Spain}

\author{Michael C. D. Tayler}
\email{michael.tayler@icfo.eu}
\affiliation{ICFO -- Institut de Ci\`encies Fot\`oniques, The Barcelona Institute of Science and Technology, 08860 Castelldefels (Barcelona), Spain}

\date{21\textsuperscript{st} June 2022}

\maketitle

\section*{KEYWORDS}
\noindent Nuclear magnetism, Composite pulses, Rotations, Low-field nuclear magnetic resonance. 

\section*{Abstract}

We discuss procedures for error-tolerant spin control in environments that permit transient, large-angle reorientation of magnetic bias field.  Short sequences of pulsed, non-resonant magnetic field pulses in a laboratory-frame meridional plane are derived.  These are shown to have band-pass excitation properties comparable to established amplitude-modulated, resonant pulses used in high, static-field magnetic resonance.  Using these meridional pulses, we demonstrate robust $z$ inversion in proton (\textsuperscript{1}H) nuclear magnetic resonance near earth's field.

\section{Introduction}

Pulsed alternating (ac) electromagnetic fields are a staple of atomic, electronic and nuclear spin resonance spectroscopies.  Following decades of development in these disciplines and others, e.g., magnetic resonance imaging\cite{Bernstein2004-book,deGraaf2019} (MRI) and quantum information processing (QIP)\cite{Jones_Les_Houches2004,Jones2011pnmrs}, there exist many species of ac pulse for precise qubit control that compensate for errors inevitably present in experimental parameters.  Among error-tolerant pulses engineered are those utilizing discrete phase-shifting\cite{Shaka1983JMR,Levitt1986pnmrs,Wimperis1994jmra,Levitt2007emr,Demeter2016}, amplitude modulation\cite{Shaka1985CPL,Shaka1987,Geen1991jmr} or both amplitude-and-phase modulation\cite{Freeman1998pnmrs,Khaneja2005jmr,Warren2007emr,Haller2022} of the ac fields. 


These pulse composition strategies are available to traditional spin-resonance experiments, which are performed inside strong magnets (e.g., superconducting magnets) with fixed magnitude and direction of the magnetic field. In this scenario, only ac amplitude and phase degrees of freedom remain for spin control.  Other strategies are in principle possible, however. At a mathematical level, error-compensated pulse design can be traced to a common set of principles, for instance the Magnus expansion\cite{Brinkmann2016cmra}, impulse-response theory\cite{Shinnar1989}, recursive iteration\cite{Levitt1983JMRrecursion} and other time symmetry considerations.  When the strong field constraint is removed, new pulse strategies become available, and existing pulse strategies can be implemented using different degrees of freedom.

In this paper we illustrate the above re-utilization concept to derive error-tolerant pulses for magnetic resonance experiments where orientation of total magnetic field is unconstrained in the laboratory frame of reference.  The case includes Earth's field nuclear magnetic resonance (NMR)\cite{Appelt2006Nphys,Callaghan2007AMR,Michal2020JMR} as well as the emerging area of zero and ultralow-field (ZULF) NMR\cite{Blanchard2016emagres,Tayler2017RSI}, which presents attractive regimes for nuclear spin hyperpolarization\cite{Sheberstov2021,VanDyke2022chemrxiv}, relaxometry\cite{Bodenstedt2021natcomm,Bodenstedt2021jpclett} and precision spectroscopy\cite{Appelt2010PRA,Blanchard2013JACS,Alcicek2021JPCLett} in fields ranging from \si{\nano\tesla} to \si{\micro\tesla}.  Here, standard ac pulses may achieve spin-species and/or transition-selective excitation\cite{SjolanderJPCA2016}.  Optimal control pulses\cite{JiangPRA2018} and direct-current (dc) analogs of ac composite pulses (e.g., 90\textsubscript{x}180\textsubscript{y}90\textsubscript{x}) \cite{ThayerJMR1986,LeeJMR1987,Bodenstedt2021jpclett} can also be used for error compensation. 

We observe that composite pulses do not always appear to translate directly from ac (high-field) to dc (low-field) techniques.  For instance, in high field, a 90\textsubscript{x}180\textsubscript{y}90\textsubscript{x} pulse is often a first choice for tolerance to error in Rabi frequency and thus pulse length.  However, in low-field, the pulse length tolerance of a dc composite pulse can be achieved using analogs of ac pulses that compensate for offset in the ac carrier frequency -- a different source of error.
This concept shall be illustrated for dc composite pulses where fields are confined to a single meridional plane of the Bloch sphere (e.g., $xz$ plane, where $z$ defines the bias axis).  We call such pulses \textit{meridional composite pulses}, and show that they are considerably more selective than traditional composite pulses, including  90\textsubscript{x}180\textsubscript{y}90\textsubscript{x} where magnetic field is kept in the equatorial plane of the Bloch sphere ($xy$ plane).

\section{Theory}

In any NMR scenario, the magnetic field $\bB(t)$ is used to produce controlled rotations of a spin $\bS$, governed by the Bloch equation
\begin{equation}
\frac{d}{dt} \bS = \gamma \bS \times \bB \,.
\end{equation}
In a high-field NMR scenario, a strong constant field  along the $z$ direction with magnitude $B_0$ is applied, and a weaker orthogonal field $B_x(t)$ is temporally shaped to produce pulses of oscillating field near the Larmor frequency $\omega_L = \gamma B_0$, with a determined detuning, duration and phase. Via the Bloch equation, such a pulse produces a spin rotation $R(\psi,\bn)$, where the spin rotation angle $\psi$ is proportional to the strength and duration of the pulse, and the (rotating frame) rotation axis $\bn$ is determined by the phase and frequency of the pulse. 

In low-field NMR, it is possible to directly implement a rotation $R(\psi,\bn)$ about a (laboratory frame) axis $\bn$, by applying a dc field of strength $B$ along $\bn$ for a time $\tau$, to generate rotation by an angle $\psi = \gamma B \tau$. Rotations with arbitrary $\bn$ and $\psi$ can in principle be produced with three-axis control of $\bB(t)$.  In this way, any simple rotation $R(\psi,\bn)$ used in high-field NMR can be implemented also in low-field NMR.

Composite pulses are not simple pulses, but rather trains of simple pulses that together implement a desired rotation.  Unlike simple pulses, these can be designed to perform nearly the same rotation for a range of parameter values, e.g., $\gamma$ or $B_0$, so these rotations become robust against experimental imperfections. They can also be used to apply different rotations to different $\gamma$ values, and thus implement species-specific rotations.  Composite pulses do not translate directly from high-field to low-field techniques, because parameter variations affect the rotation in different ways. For example, in a resonant rotation, $\bn$ depends on the detuning and thus on $\gamma$, whereas for a dc rotation it is $\psi$ that depends on $\gamma$. 

As a starting point for meridional composite pulse design we use the theorem that successive rotation of a spin (and more generally, any 3d object) by $\pi$ radians about an arbitrary pair of unit vectors $\bn'$ and $\bn''$ is equivalent to a single rotation by an angle $\phi$ about the perpendicular unit vector $\bn \propto \bn' \times \bn''$, where $\phi/2$ is the angle between $\bn'$ and $\bn''$, i.e. $\bn' \cdot \bn'' = |\bn'| |\bn''| \cos (\phi/2)$:
\begin{equation}
\label{eq:rr}
	R(\pi, \bn'') R(\pi, \bn') = R(\phi, \bn)\,.
\end{equation} 
By extension, an equation follows for the cumulative effect of $2N$ rotations-by-$\pi$ about axes $\bn_1', \bn_1''$, $\bn_2', \bn_2'', \ldots, \bn_N', \mathbf{n}_N''$ in a common plane normal to $\bn$:
\begin{equation}
	\prod_{j=1}^N R(\pi, \bn_{j}'') R(\pi,\bn_{j}') \equiv R(\sum_{j=1}^N \phi_j, \bn)\,. \label{eq:product1}
\end{equation}
For instance, if all of the $\bn'$ and $\bn''$ vectors lie within the Cartesian $xz$ plane as defined by vectors $\bx = (1,0,0)$ and $\bz = (0,0,1)$, then the overall rotation is produced about the $y$ axis, defined $\by = (0,1,0)$; $\bn=\by$. 

The problem of interest for robust, spin-selective pulse generation is the approximate implementation of \autoref{eq:product1}, where a sequence of $2N$ rotations by $\kappa\pi$ is applied about axes $\bn'_1, \bn''_1, \ldots , \bn'_N, \bn''_N$.  If, as above, the angle between $\bn'_i$ and $\bn''_i$ is $\phi_i/2$, the objective is to find a sequence of angles $\phi_1, \phi_2, \ldots , \phi_N$ such that the resulting rotation is
\begin{equation}
\tilde{R}(\kappa) \equiv 
\prod_{j=1}^N R(\kappa\pi, \bn_{j}'') R(\kappa\pi,\bn_{j}') \approx R(\beta,\bn) \label{eq:product2}
\end{equation}
for some detuned range of $\kappa$, (and therefore gyromagnetic ratio, $\gamma\propto\kappa$), say ($\kappa$ mod $2$) = $(1 + \delta)$, where $\delta$ is the detuning.  Here  $\beta$ is the target rotation angle.  




One route to a solution is to recognize that for $\kappa=(1+\delta)$ one can cast \autoref{eq:product2} (see \hyperref[appx:Eq4derivation]{Appendix A}) into a form
\begin{eqnarray}
\tilde{R}(\kappa) &=&\prod_{j=1}^N  R(\frac{\phi_j}{2},\by) R(\pi[1+\delta], \bz) R(-\frac{\phi_j}{2},\by) R(\pi[1+\delta],\bz) \nonumber \\
	&=& R_z^{2N} \prod_{j=1}^N 
	\bigl[\hat{R}_z^{2j}(\delta)R(\frac{\phi_j}{2},\by)\bigr] 
	\bigl[\hat{R}_z^{2j-1}(\delta)R(-\frac{\phi_j}{2},\by)\bigr],
	\nonumber \\
\label{eq:rotatingproduct}
\end{eqnarray}
where $\hat{R}_z^d(\delta)$ is the $d$th power of the right-acting superoperator $\hat{R}_z(\delta)$, which rotates operators $R(\phi,\bn)$ by an angle $\pi(1+\delta)$ about $\mathbf{z}$, as defined by $\hat{R}_z(\delta)R(\phi,\bn) \equiv R(-\pi[1+\delta],\bz) R(\phi,\bn) R(\pi[1+\delta],\bz)$ \cite{JEENER1982}.  The form of \autoref{eq:rotatingproduct} indicates that, relative to the ideal transformation $(\kappa=1)$, the error $\delta$ has the effect of shifting the spins' frame of reference by an offset $2\delta$ about $\bz$ between each $\phi_j$ rotation.  In this way, the problem of finding a suitable set of $\phi_j$s is mapped onto another problem; that of compensating for frame offset.   
Frame-offset compensation is a well-explored topic in physics, and of high importance in NMR, MRI and QIP.  One representative strategy uses {broadband uniform-rotation pure-phase} (BURP\cite{Geen1991jmr}) pulses, as first developed by Geen and Freeman\cite{Freeman1998pnmrs}. A BURP pulse is an amplitude-modulated ac pulse of duration $\tau_p$, with the carrier resonant with the nominal Larmor frequency. The carrier envelope is chosen such that for detuning $\delta = 0$, the (rotating frame) rotation is $R(\beta(t),\by)$, where for $0 \le t \le \tau_p$ the accrued flip angle is 
\begin{equation}
    \beta(t) = a_0 \frac{t}{\tau_p} + b_0 + \sum_{k=1}^{n_\mathrm{cut}}{ \left[ a_k \sin\left(\frac{2\pi k t}{\tau_p}\right) + b_k \cos\left(\frac{2\pi k t}{\tau_p}\right) \right]}.  
    \label{eq:BURPdefinition}
\end{equation}
This is a truncated Fourier series, and a cutoff of $n_\mathrm{cut} \sim 6$ typically gives sufficient precision \cite{Geen1991jmr,Freeman1998pnmrs};  numerical values for $a_k$ and $b_k$ are given in the Supplemental Material\cite{SM}.  BURP pulses generously tolerate mismatch between the carrier and Larmor precession frequencies, with an excitation pass-band inversely proportional to the BURP pulse length.

Using dc pulse pairs as described in \autoref{eq:product1}, we can make a pointwise approximation of $\beta(t)$: We define intermediate rotation angles 
\begin{equation}
    \phi_j = \beta\left(\frac{\tau_p j }{N}\right) - \beta\left(\frac{\tau_p (j-1) }{N}\right)\,,
\end{equation}
and then construct a sequence of nominally-$\pi$ rotations 
\begin{equation}
\tilde{R}(\kappa) \equiv 
\prod_{j=1}^N R(\kappa\pi, \bn_{j}'') R(\kappa\pi,\bn_{j}') 
\label{eq:productBURP}
\end{equation}
with $\bn_j'=(+\sin(\phi_j/4),\,0,\,\cos(\phi_j/4))$ and $\bn_j''=(-\sin(\phi_j/4),\,0,\,\cos(\phi_j/4))$. This defines a \textit{meridional composite pulse}, i.e., a series of $\kappa\pi$ rotations about pairs of axes in the $xz$ plane, separated by angles $\phi_j/2$\footnote{While the BURP pulse has a duration $\tau_p$, the pointwise approximation does not.  This is because the actual rotation sequence does not depend on absolute value of $\tau_p$, as illustrated by \autoref{eq:BURPdefinition}.}. 

\begin{figure}
    \centering
    \includegraphics[width=0.9\columnwidth]{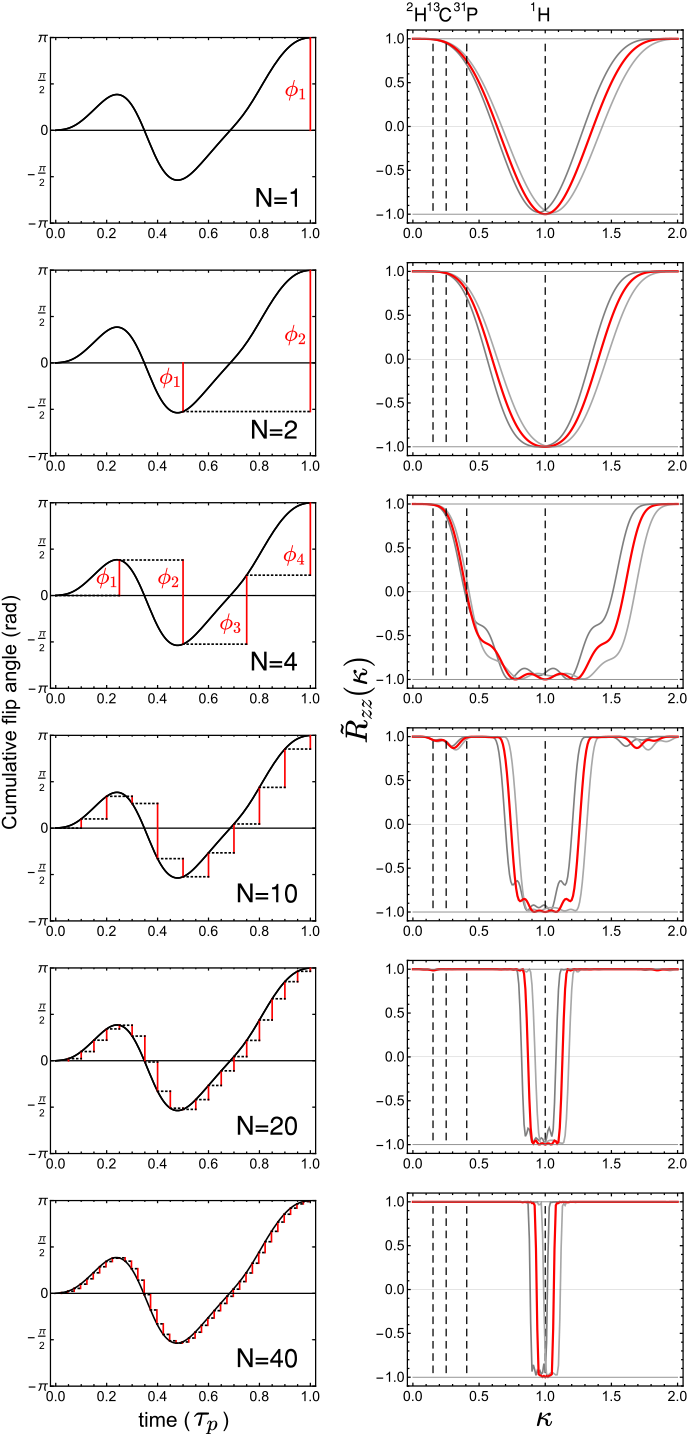}
    \caption{Band-pass spin inversion using dc pulses derived from I-BURP-1\cite{Geen1991jmr}.  The left column shows time-profiles of flip angle accumulated in the $xz$ plane for the conventional ($\beta$, black) and discretized ($\phi_j$, red) pulses with $N=1,2,4,10,20,40$.  Plots to the right show $z$ inversion performance $\tilde{R}_{zz}(\kappa)$ for the discretized pulse, using scaling factors $s=1$ (red), $s=0.9$ (gray), $s=1.1$ (light gray) as defined in \autoref{eq:Bvector}.  Dashed vertical lines show values of $\kappa(S)/\kappa($\textsuperscript{1}H$) \equiv \gamma_S/\gamma_H$, for the case $\kappa$(\textsuperscript{1}H) = 1.}
    \label{fig:IBURP}
\end{figure}

Performance of the discretized BURP pulses can be analyzed by numerical simulation.  As a first example, we study the inversion pulse I-BURP-1\cite{Geen1991jmr}, as shown in \autoref{fig:IBURP}.  Continuous and pointwise $\beta(t)$ values agree closely with one another for $N \gtrsim n_{\rm max}$. For instance, for $N>10$, the fractional difference between $\beta(t)$ and the connecting line between sampling points $\beta(j\tau_p/N)$ is below $0.05$ for all time points.

We quantify the inversion by $\tilde{R}_{zz}(\kappa) \equiv \bz^T \tilde{\vb*{R}}(\kappa) \bz$, where  $\tilde{\vb*{R}}(\kappa)$ is the matrix representation of the net rotation operator $\tilde{R}(\kappa)$.  A value $\tilde{R}_{zz}(\kappa)=-1$ implies complete spin inversion, while $\tilde{R}_{zz}(\kappa)=+1$ indicates zero net rotation of the spin away from $\bz$. 
From plots on the right side of \autoref{fig:IBURP}, $\tilde{R}_{zz}(\kappa)$ shows an inversion passband of full width at half-maximum $2\delta \approx 5/N$, which for moderate values $N\sim 20$ should be wide enough to provide generous error tolerance, e.g., $2\delta \sim 0.25$, while being selective in $\gamma$. 

DC field pulses are typically produced by field coils, with each coil contributing the field component along one Cartesian axis.  To implement the pulse sequence of \autoref{eq:rotatingproduct}, for example, X (field along $\bx$) and Z (field along $\bz$) coils could be used. We now analyze the effect of a mis-calibration of the Z coil by a factor $s$, so that the produced field is 
\begin{equation}
    \va{B} = B_0 (\sin\varphi,0,s\cos\varphi) \,,\label{eq:Bvector}
\end{equation}
where $B_0$ is the intended field strength and $\varphi$ is the intended angle in the $xz$ plane. For $s \approx 1$, the effect on the rotation angle, i.e.\ on $\delta$, is first order in $s-1$, and the effect on $\bn$ is a non-simple function of $\varphi$ and $\phi$.  For symmetric displacement of $\bn'$ and $\bn''$ about $\bz$ (and thus $\bn'_j+\bn''_j$ parallel to $\bz$, when $\varphi=\pm\phi/2$) the excitation profile remains mostly unchanged with respect to $s$ and the passband center shifts to $\kappa\approx(1+s)/2$.  Representative profiles for the I-BURP-1 pulse are shown in \autoref{fig:IBURP}.

Another result of this approach is dc analogs of wide-offset-tolerant ac composite pulses.  These pulses in high-field NMR are often termed ``phase-alternating composite pulses'' \cite{Shaka1985CPL,Shaka1987,Yang1995,RamamoorthyMP1998,Husain2013} due to the alternating sign of flip angle in the ac frame, e.g., $(\beta_1,\,\beta_2,\,\beta_3)$ = (\SI{59}{\degree},\,\SI{-298}{\degree},\,\SI{59}{\degree})\cite{Shaka1987}.  The flip angles can be directly mapped to a dc meridional composite pulse using $\phi_j=\beta_j$.

\begin{table*}[t]
	\centering
\begin{tabular}{ccccccccccccc}
	\hline \hline
$N$ &  \multicolumn{9}{c}{angles (degree)} & Stopband & Passband & Reference \\
 & $ \phi_1 $ & $ \phi_2 $ & $ \phi_3 $ & $ \phi_4 $ & $ \phi_5 $ & $ \phi_6 $ & $ \phi_7 $ & $ \phi_8 $ & $ \phi_9 $ &   \quad ($\tilde{R}_{zz}(\kappa)>0.99$) \quad & \quad ($\tilde{R}_{zz}(\kappa) <-0.95$) \quad &  \\\hline  
2	& 55 &-235 &&&&&&&& $|\kappa|<0.20$ & $0.83<\kappa<1.17$ & this work    \\ 
3	& 59& -298& 59 	 &&&&&&& $|\kappa|<0.36$ & $0.93<\kappa<1.07$ & Shaka \etal\cite{Shaka1987}    \\  
3	& 24& -97& 253  &&&&&&& $|\kappa|<0.24$ & $0.75<\kappa<1.25$ 	& this work     \\  
4	& -34&  123&  -198&  289 	  &&&&&& $|\kappa|<0.29$ & $0.86<\kappa<1.14$ & Shaka 
\cite{Shaka1985CPL}, Yang \etal\cite{Yang1995}    \\
4	& 27& -81& 263& -30 &&&&&&  $|\kappa|<0.29$  & $0.75<\kappa<1.25$	& this work    \\  
5 & 325&  -263&  56&  -263&  325 &&&&& $|\kappa|<0.41$ & $0.95<\kappa<1.05$ & Shaka \etal\cite{Shaka1987} \\
9 &70&  -238&  -355&  296&  276&  296&  -355&  -238&  70	& $|\kappa|<0.55$	& $0.89<\kappa<1.11$	 & this work  \\ \hline \hline
\end{tabular}
\caption{Composite pulses for the inversion operation $\mathbf{z}\rightarrow-\mathbf{z}$.  All angles $\phi_j$ are given in degrees.} 
\label{tab:pulselist}
\end{table*}

\begin{figure}[h]
    \centering
    \includegraphics[width=0.9\columnwidth]{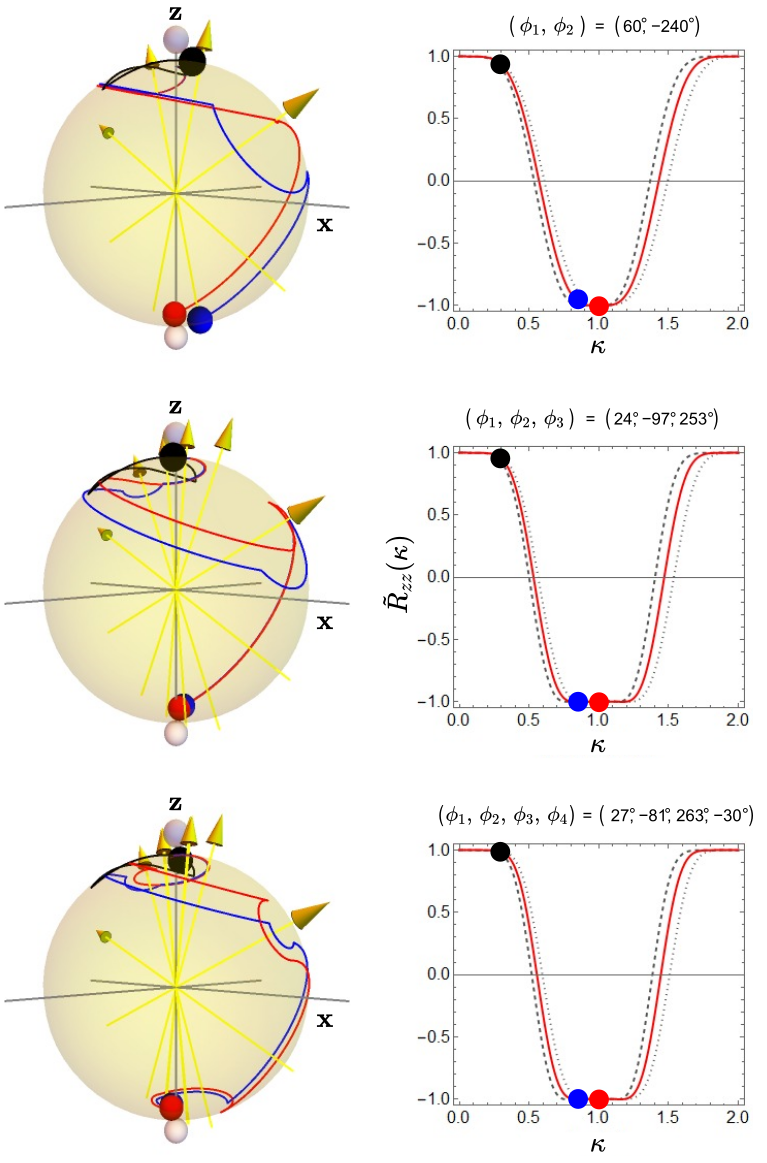}
    \caption{Spin vector trajectories in the Bloch sphere and band-pass inversion profiles for selected composite pulses listed in \autoref{tab:pulselist}.  The black, blue and red trajectories are for values $\kappa=0.3$, $\kappa=0.85$ and $\kappa=1.0$.  Rotation axes indicated by yellow arrows are displaced symmetrically about $\bz$ in the $xz$ plane.  The solid red curve indicates the band-pass profile $\bz\rightarrow-\bz$ for $s=1$; dashed and dotted gray curves correspond to profiles for $s=0.9$ and $s=1.1$, respectively. }
    \label{fig:fig2}
\end{figure}

Selected phase-alternating composite pulses for inversion and the widths of their passbands are listed in \autoref{tab:pulselist}.  Highly uniform inversion can be achieved using only a few pulses ($N<10$), with a degree of selectivity comparable to, if not better than I-BURP-1.  
We note that the passband widths vary between the works of different authors because of different optimization criteria.  Uniform excitation within the passband is often given the highest priority, followed by rejection in stopband.  Because in some low-field NMR applications both figures of merit may have equal priority, the present work includes some additional solutions. 
The original sequences we report in \autoref{tab:pulselist} are found in a few minutes with a standard desktop computer, by randomly sampling $\sim 50000$ points in a $N-1$ dimensional space of $\phi$ values $\phi_1, \phi_2,\,\ldots ,\phi_{N-1}$, with resolution $0.02\times\pi$ \SI{}{\radian}. The final angle is constrained to be $\phi_N=\pi-\sum_{j=1}^{N-1}\phi_j$, so that $\beta=\pi$. Our merit function is $l_1+l_2$, where $l_1$ is the mean value of $(1 +  \tilde{R}_{zz}(\kappa))^2$ over the range $0.8<\kappa<1$ and $l_2$ is the mean of $(1 - \tilde{R}_{zz}(\kappa))^2$ for $|\kappa|<0.5$.  
Angle sets up to length $N=9$ giving widest passband and stopband widths are presented in \autoref{tab:pulselist}.  Generally, we observe these sequences can have a wider passband than the existing phase-alternating composite pulses of the same length $N$.  An increased width of the stopband is more challenging and requires higher $N$.  The performance of these pulses is illustrated in \autoref{fig:fig2}.

\section{Experimental results}

The band-pass profiles of meridional composite pulses such as those shown in \autoref{fig:IBURP} and \autoref{fig:fig2} can be measured using a sample containing only a single spin species, e.g.\ \textsuperscript{1}H in water (\textsuperscript{1}H\textsubscript{2}O).  We note that the rotation axes $\bn'$ and $\bn''$ are independent of $\gamma$, while $\kappa$ is directly proportional to $\gamma$ and pulse duration  Thus the effects of a change in $\gamma$ can be simulated by a corresponding change in the pulse duration.  

\begin{figure}
    \centering
    \includegraphics[width=\columnwidth]{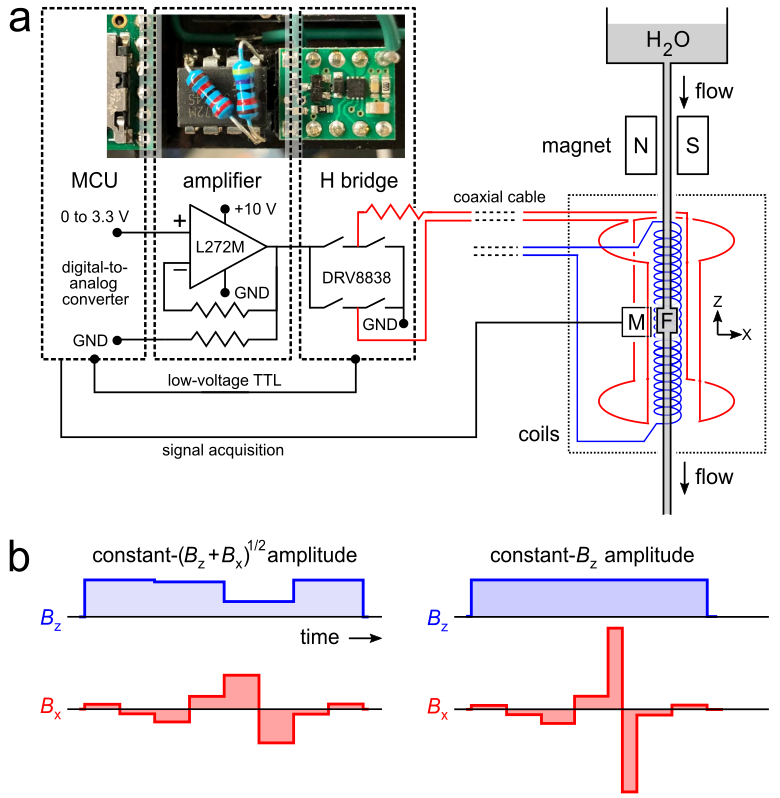}
    \caption{Experimental setup to test dc meridional composite pulses. (a) NMR detection apparatus and electrical circuit used to supply variable current to X and Z field coils.  Typical field switching and settling times are $<$\SI{2}{\micro\second}.  M = alkali-atom vapor magnetometer used to detect NMR signals.  F = Flow cell containing spin-polarized water, MCU = Microcontroller Unit. TTL = 3.3 V transistor-transistor logic signal; (b) Time profile of dc field amplitude for the $N=4$ sequence $(\phi_1,\,\phi_2,\,\phi_3,\,\phi_4) = (\SI{27}{\degree},\SI{-81}{\degree},\SI{263}{\degree},\SI{-30}{\degree})$, where magnetic fields are applied in the $xz$ plane.  Two possible implementations are shown.  On the left, total field amplitude $B_0$ and pulse duration are kept constant during each pulse section, while on the right, the pulse lengths and amplitudes are scaled for constant $B_z$.  
    }
    \label{fig:SetupAndBxBzSequences}
\end{figure}

The experimental setup and testing protocol are shown in \autoref{fig:SetupAndBxBzSequences}a.  Water, pre-polarized along $z$, flows through a cell surrounded by two coils (X and Z) that produce uniform fields along $\bx$ and $\bz$, respectively.  Bipolar current control of the X coil is provided by a simple electronic circuit comprising a digital-to-analog converter, amplifier and H-bridge module. A constant current is passed through the Z coil, as in \autoref{fig:SetupAndBxBzSequences}b (right). In this arrangement (unlike what is suggested by \autoref{eq:Bvector}), the field strength $B_0$ depends on the angle $\varphi$. The pulse duration $\tau$ is compensated accordingly, so that the nominal rotation angle $\psi = \gamma B_0 \tau$ is always $\pi$. Further details are given in the Methods section.

The performance of the spin-selective inversion pulses is measured by applying a composite pulse, then immediately applying a dc pulse of flip angle \SI{90}{\degree} along $+\bx$.  The peak field of the composite pulse is controlled such that I-BURP-1 pulse lengths are of comparable length for different $N$: $(\tau_p / \kappa) \sim \SI{8}{\milli\second}$ for \textsuperscript{1}H.  An alkali-metal-vapor magnetometer\cite{Bodenstedt2021natcomm} adjacent to the flow cell detects the resulting \textsuperscript{1}H free precession signal (FID).  

\begin{figure*}
    \centering
    \includegraphics[width=0.95\textwidth]{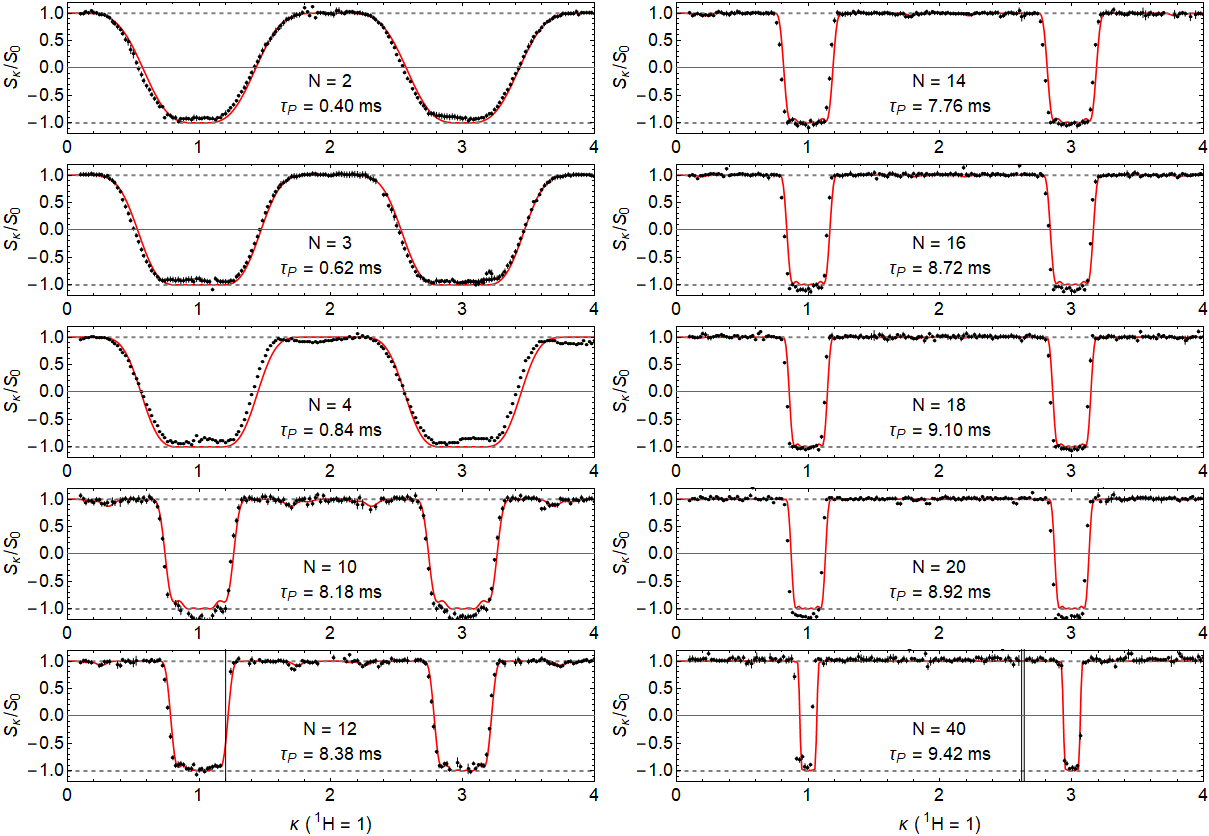}
    \caption{Experimental (black circles) vs.\ simulated (solid red curves) band-pass inversion profiles for meridional composite pulses with $N$ = 2, 3, 4 listed as ``this work'' in \autoref{tab:pulselist} and for discrete I-BURP-1 pulses with $N$ = 10, 12, 14, 16, 18, 20 and 40.  Each data point represents the mean signal amplitude of approximately 10 NMR transients.}
    \label{fig:fig4}
\end{figure*}

The observed FID amplitude for composite pulses of duration $\kappa\tau_p$ is denoted $S_{\kappa}$, and the FID amplitude with no applied composite pulse is $S_{0}$.  The ratio $S_\kappa/S_0$ equals $\tilde{R}_{zz}(\kappa)$, which takes values between $-1$ (for complete spin inversion) and $+1$ (for no spin inversion) and is plotted up to $\kappa=4$ for various pulses in \autoref{fig:fig4}.

The experimental and simulated profiles agree closely, with residuals below the experimental error margins.  This result confirms that spin selective pulses can indeed be designed using the approach of \autoref{eq:product1}.  It also suggests that any imperfections in the pulses are small compared to the compensation limits, which is remarkable considering the simplicity of the electronic drive circuitry.

\section{Discussion and conclusions}

It is shown theoretically and confirmed by experiment that efficient, robust, and spin-selective composite pulses can be composed using switched dc fields in a meridional plane.  Also identified is an equivalence between the problems of flip-angle tolerance in low-field NMR, and the well-studied problem of frame rotation tolerance in high-field NMR. 

In contrast to high-field NMR, where phase/frequency offsets and flip-angle offsets originate in distinct imperfections of the experimental system and have different effects on the generated spin rotations, in low-field NMR flip angle and axis are both determined by the dc field strengths.  As shown, this allows a single composite pulse strategy to be robust against variations of each, something that is uncommon in high-field composite pulses.

Also unlike many high-field NMR composite pulses, the criterion determining total flip angle of a meridional composite pulse (\autoref{eq:product1}) is not limited to special angles (e.g., \SI{90}{\degree}, \SI{180}{\degree}) and thus should allow one to obtain pulses of arbitrary flip angle.  This general design approach, coupled with the above error compensation properties, should prove valuable in spin resonance applications at low field that require robust and selective control.  The pulses demonstrated here are much shorter in duration than high-frequency ac pulses of equivalent compensation bandwidth, such as swept-frequency adiabatic inversion pulses\cite{Tayler2016JMR}, and can be performed without tuned high-frequency circuitry.  
Application is expected in sub-MHz NMR spectroscopy and MRI, field-cycling relaxation measurements, nuclear spin polarimetry, as well as portable NMR spectrometers for use outside of the research laboratory.

The pulse durations $\tau_p$ in the present study are limited by hardware timer resolution (\SI{2}{\micro\second}) and the field-to-current ratio of the X coil.  Faster clock speeds and stronger fields could shorten pulse lengths by at least one order of magnitude, giving $\tau_p$ between $\SI{10}{\micro\second}$ and $\SI{100}{\micro\second}$.   These durations are much shorter than the periods of the spin-spin scalar couplings between common nuclear spin species, and thus should be applicable to heteronuclear quantum control in low-field NMR, in which pulses selectively rotate one or more spins in a multi-species system.   The selectivity and error tolerance should be complementary to existing control methods based on equatorial composite pulses \cite{Bian2017physrevA, Bodenstedt2021jpclett}.


\section*{Methods}

Experimental testing of the composite pulses utilized a continuous-flow (a.k.a.\ ``polarization on tap'') test sample for high-throughput measurement.  As shown in \autoref{fig:SetupAndBxBzSequences}a, distilled water from a reservoir of several liters capacity drained continuously under gravity (flow rate $\sim$\SI{1}{\milli\liter\per\second}) through a low-homogeneity 1.5 T magnet\cite{Tayler2017JMR} allowing the \textsuperscript{1}H spins to reach thermal equilibrium polarization.  The liquid subsequently flowed into the $\sim$\SI{1}{\milli\liter} sample chamber,
with the sample magnetization of around \SI{1}{\pico\tesla} being aligned parallel to the axis of the background field, along $\bz$. 

Centered on the sample chamber was a solenoid coil ($\sim$7.5 mT/A) and a saddle coil ($\sim$\SI{80}{\micro\tesla/A}) to produce magnetic fields along $\bz$ and $\bx$, respectively.  Currents applied to the saddle coil were controlled using a simple dc switch comprising a microcontroller (ARM Cortex M4F) digital-to-analog converter (12 bits, 0 to 3.3 V), operational amplifier (L272M, STMicroelectronics) and an H-bridge module (Texas Instruments DRV8838 on Pololu 2990 carrier board), as shown in \autoref{fig:SetupAndBxBzSequences}a.  Rabi curves were measured for different values of the DAC output to confirm a linear output of voltage across the coils between 0.4 and 10 V (See Supplemental Material\cite{SM}).  This determined the switchable range of field for a given series resistance of the coil.  The 10 V and ground op-amp rails were connected to a standard laboratory power supply unit (Hameg HM7042-5).

Details of magnetometer used to measure the NMR signals can be found in previous work\cite{Bodenstedt2021natcomm}.  The signals $S_\kappa$ and $S_0$ of the two pulse sequences are acquired in an interleaved fashion to minimize the effect of drifts. 

\section*{Acknowledgement}
The work described is funded by: 
EU H2020 Marie Sk{\l}odowska-Curie Actions project ITN ZULF-NMR  (Grant Agreement No. 766402); 
Spanish MINECO project OCARINA (PGC2018-097056-B-I00 project funded by MCIN/ AEI /10.13039/501100011033/ FEDER “A way to make Europe”); 
the Severo Ochoa program (Grant No. SEV-2015-0522); 
Generalitat de Catalunya through the CERCA program; 
Ag\`{e}ncia de Gesti\'{o} d'Ajuts Universitaris i de Recerca Grant No. 2017-SGR-1354;  Secretaria d'Universitats i Recerca del Departament d'Empresa i Coneixement de la Generalitat de Catalunya, co-funded by the European Union Regional Development Fund within the ERDF Operational Program of Catalunya (project QuantumCat, ref. 001-P-001644);
Fundaci\'{o} Privada Cellex; 
Fundaci\'{o} Mir-Puig; 
MCD Tayler acknowledges financial support through the Junior Leader Postdoctoral Fellowship Programme from ``La Caixa'' Banking Foundation (project LCF/BQ/PI19/11690021).


\section*{Author Contributions}
MCD Tayler made the theoretical interpretation and wrote the manuscript with input from all authors.  
S. Bodenstedt built the experimental apparatus, measured and analyzed the experimental data. 
MCD Tayler and MW Mitchell supervised the overall research effort. 

\section*{Competing Interests}
There are no competing interests to declare.

\appendix
\section{Derivation of Eq. (5) } 

\label{appx:Eq4derivation} 
Starting from \autoref{eq:product2}, we choose rotation axes $\bn'_j=\bz$ and $\bn''_j=(\sin[\phi_j/2],\,0,\,\cos[\phi_j/2])$ that subtend angle $\phi_j/2$ and then use the Euler YZY convention to rewrite the product in terms of rotations about the Cartesian $\bz$ and $\by$ axes:
\begin{equation}
    \tilde{R}(\beta,\bn) \,\,=\,\, \prod_{j=1}^N R(\frac{\phi_j}{2},\by)R(\pi[1+\delta],\bz)R(-\frac{\phi_j}{2},\by) R(\pi[1+\delta],\bz)\,.\label{eq:yzy}
\end{equation}
Then by using $y_j\equiv R(\phi_j/2,\by)$, $z\equiv R(\pi[1+\delta],\bz)$, and an overbar to denote opposite sign of $\phi$, the product in \autoref{eq:yzy} can be written in a shorthand notation
\begin{equation}
     \tilde{R}(\beta,\bn) \,\,=\,\, \prod_{j=1}^N  (y_j z \overline{y}_j) z \,\, \equiv \,\, y_N z \overline{y}_N \ldots z y_2 z \overline{y}_2 z y_1 z \overline{y}_1 z\,. \label{eq:yzyshorthand}
\end{equation}
The next step is to insert the identity operation $1 \equiv z^{j} z^{-j}$ in between every $zy$ or $z\overline{y}$ product, in order to obtain products of the form $z^{-j}yz^{j}$.  The right side of \autoref{eq:yzyshorthand} then equals
\begin{equation}
    z^{2N}(z^{-2N} y_N z^{2N}) (z^{2N-1} \overline{y}_N z^{2N-1}) \ldots (z^{-2} y_1 z^2) (z^{-1}\overline{y}_1 z)\,,
\end{equation}
and therefore also equals the right side of \autoref{eq:rotatingproduct} when writing back in long form, using $z^{-i}y_jz^{i} = \hat{R}_z^iR(\phi_j/2,\by)$, $z^{-i}\overline{y}_jz^{i} = \hat{R}_z^iR(-\phi_j/2,\by)$ and so on.



\bibliography{references}

\end{document}